\documentclass[twocolumn]{aastex62}

\usepackage{amsmath}

\def\kms{km~s$^{-1}$}

\graphicspath{{./}{figures/}}

\received{August 1, 2018}
\revised{August 24, 2018}
\accepted{August 28, 2018}

\submitjournal{ApJL}

\shorttitle{Pulsar Wind Nebula Candidate in SN\,2012au}
\shortauthors{Milisavljevic et al.}

\begin{document}

\title{Evidence for a pulsar wind nebula in the Type Ib-peculiar supernova SN\,2012au}

\correspondingauthor{Dan Milisavljevic}
\email{dmilisav@purdue.edu}

\author[0000-0002-0786-7307]{Dan Milisavljevic}
\affil{Department of Physics and Astronomy,
Purdue University, 525 Northwestern Ave.,
West Lafayette, IN. 47907, USA}

\author[0000-0002-7507-8115]{Daniel J.\ Patnaude}
\affiliation{Smithsonian Astrophysical Observatory, 60 Garden St., Cambridge, MA 02138, USA}

\author[0000-0002-9117-7244]{Roger A.\ Chevalier}
\affiliation{Department of Astronomy, University
of Virginia, P.O.\ Box 400325, Charlottesville, VA
22904, USA}

\author[0000-0002-7868-1622]{John C.\ Raymond}
\affiliation{Smithsonian Astrophysical Observatory, 60 Garden St., Cambridge, MA 02138, USA}

\author[0000-0003-3829-2056]{Robert A.\ Fesen}
\affiliation{6127 Wilder Laboratory, Department of Physics and Astronomy, Dartmouth College, Hanover, NH 03755, USA}

\author[0000-0003-4768-7586]{Raffaella Margutti}
\affiliation{Center for Interdisciplinary Exploration and Research in Astrophysics (CIERA) and Department of Physics and Astronomy, Northwestern University, Evanston, IL 60208}

\author{Brody Conner}
\affil{Department of Physics and Astronomy,
Purdue University, 525 Northwestern Ave.,
West Lafayette, IN. 47907, USA}

\author{John Banovetz}
\affil{Department of Physics and Astronomy,
Purdue University, 525 Northwestern Ave.,
West Lafayette, IN. 47907, USA}

\begin{abstract}

We present an optical spectrum of the energetic Type Ib supernova (SN) 2012au obtained at an unprecedented epoch of 6.2 years after explosion.  Forbidden transition emission lines of oxygen and sulfur are detected with expansion velocities of $\approx 2300$ \kms. The lack of narrow H Balmer lines suggests that interaction with circumstellar material is not a dominant source of the observed late-time emission. \added{We also present a deep {\em Chandra} observation that reveals no X-ray emission down to a luminosity of $L_{X} < 2\times 10^{38}$\,erg\,s$^{-1}$ (0.5--10\, \rm keV).} Our findings are consistent with the notion that SN~2012au is associated with a diverse subset of SNe, including long-duration gamma-ray burst SNe and superluminous SNe, harboring pulsar/magnetar wind nebulae that influence core-collapse explosion dynamics on a wide range of energy scales. We hypothesize that these systems may all evolve into a similar late-time phase dominated by forbidden oxygen transitions, and predict that emission line widths should remain constant or broaden a few per cent per year due to the acceleration of ejecta by the pulsar/magnetar bubble. 

\end{abstract}


\section{Introduction} \label{sec:intro}

Models of hydrogen-poor and energetic ($E_k \sim 10^{52}$ erg) broad-lined Type Ic (Ic-bl)  core-collapse supernovae (SNe) associated with long-duration gamma-ray bursts (GRBs) often invoke central engine-driven mechanisms associated with the formation of compact objects that input energy into the explosion. The ``collapsar'' mechanism \citep{MacFadyen99}, which involves a massive star collapsing directly to a black hole with accretion releasing energy in the form of a relativistic jet that can explode the star, has long been a favored model \citep{WH12}. However, mounting evidence supports the view that these explosions may be powered by rapidly rotating ``magnetars,'' i.e., a neutron star with an exceptionally strong ($> 10^{14}$ G) magnetic field \citep{Thompson04,Metzger11,Mazzali14,Metzger15}. 

\begin{figure*}
\centering
\includegraphics[width=0.9\linewidth]{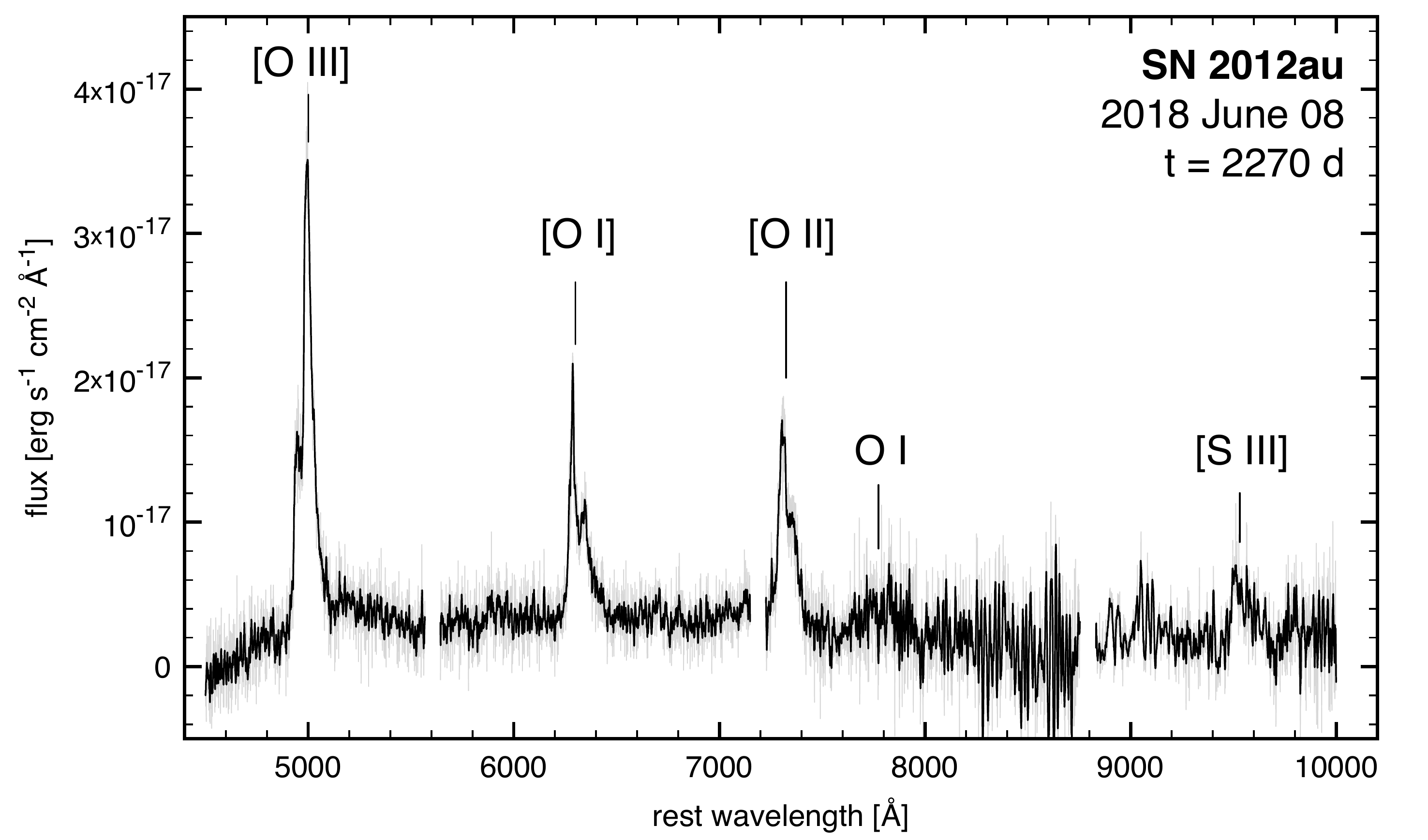}\\
\includegraphics[width=0.85\linewidth]{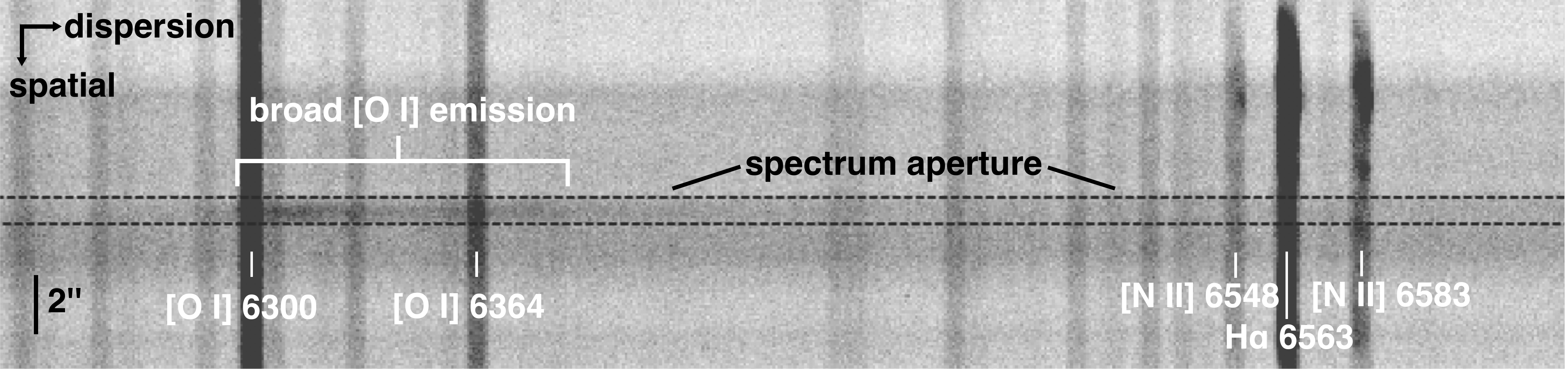}

\caption{Magellan+IMACS optical spectrum of SN\,2012au obtained 2018\,June\,8. {\it Top}: Entire 1D spectrum as extracted (gray) and data smoothed using 5~\AA\ boxcar (black). The phase is with respect to the $v$-band maximum on 2012 March 21. \added{{\it Bottom}: 2D spectrum in the region of [\ion{O}{1}] and H$\alpha$. No conspicuous narrow emission lines local to the SN are observed.}}

\label{fig:fullspec}
\end{figure*}

Magnetars have also been used in models characterizing the diverse class of hydrogen-poor superluminous SNe (\added{SLSNe-I}; \citealt{Inserra13,Nicholl13}). SLSNe-I can radiate more than $10^{44}$\,erg\,s$^{-1}$ at their peak luminosity and may be associated with extremely massive ($> 100$ M$_{\odot}$) progenitor stars. Their slow-evolving light curves are consistent with those expected from the decay of tremendous amounts of radioactive $^{56}$Ni ($> 3$ M$_{\odot}$), as might be synthesized by a pair-instability explosion. However, this scenario fails in numerous cases where the luminosity reaches levels requiring the nickel mass to be larger than the estimated ejecta mass. 

Interaction between a SLSN-I and surrounding circumstellar material (CSM) is another possible energy source \citep{Chatzopoulos12}. However, SLSN-I events generally lack conspicuous spectroscopic features that support this interpretation at photospheric stages. SLSN--CSM interaction also requires an extreme mass-loss history of several M$_{\odot}$ of H-poor material shed in the last year before explosion in order to reproduce the observed light curves \citep{CI11,Lunnan18}. 

Discovery of SN\,2011kl, which had a higher-than-average luminosity and was associated with the ultra-long-duration GRB\,111209A, suggests that GRB and SLSN classifications are not necessarily distinct \citep{Greiner15}. Understanding the SLSN--GRB connection is presently an area of active investigation \citep{PM17,Coppejans18,Margutti18}. The connection was anticipated by earlier observations of the SLSN 2010gx \citep{Pastorello10}, and the unusually energetic ($E_k\sim 10^{52}$\,erg), slow-evolving, Type Ib SN\,2012au, which is believed to be a lower-luminosity counterpart to SLSNe  \citep{Milisavljevic13,Kamble14,MM18}. Nebular phase observations of the SLSN\,2015bn \citep{Nicholl16} \added{and other SLSNe-I \citep{Nicholl18}} have further strengthened the connection. Core angular momentum may be the key ingredient differentiating SLSNe and GRBs \citep{Lunnan14}. However, other factors including star formation rate and stellar mass may also be influential \citep{Angus16}. Recent speculation that Fast Radio Bursts (FRBs) may be magnetars in the remnants of SLSN explosions has heightened the significance of the SLSN--GRB connection \citep{Metzger17,Nicholl17,Eftekhari18}. 

Here we present optical and X-ray data of SN\,2012au obtained $> 6$ yr post-explosion that offer fresh insight into the suspected link between SLSNe and GRB-SNe. In sections \ref{sec:observations} and \ref{sec:results} we present our optical spectroscopy and supporting X-ray observations, along with analysis and results. In sections \ref{sec:discussion} and \ref{sec:x-ray} we discuss possible excitation mechanisms and outline numerous arguments in favor of heating of ejecta by a pulsar wind nebula (PWN). Our discovery marks the first time a PWN signature has been detected in a verified extragalactic SN~Ib at this extremely late epoch. We conclude in section \ref{sec:conclusions} with implications of our result and suggest future avenues of investigation. We adopt $23.5 \pm 0.5$ Mpc as the distance to the host galaxy NGC 4790 \citep{Theureau07}. 

\begin{figure*}
\centering
\includegraphics[width=0.45\linewidth]{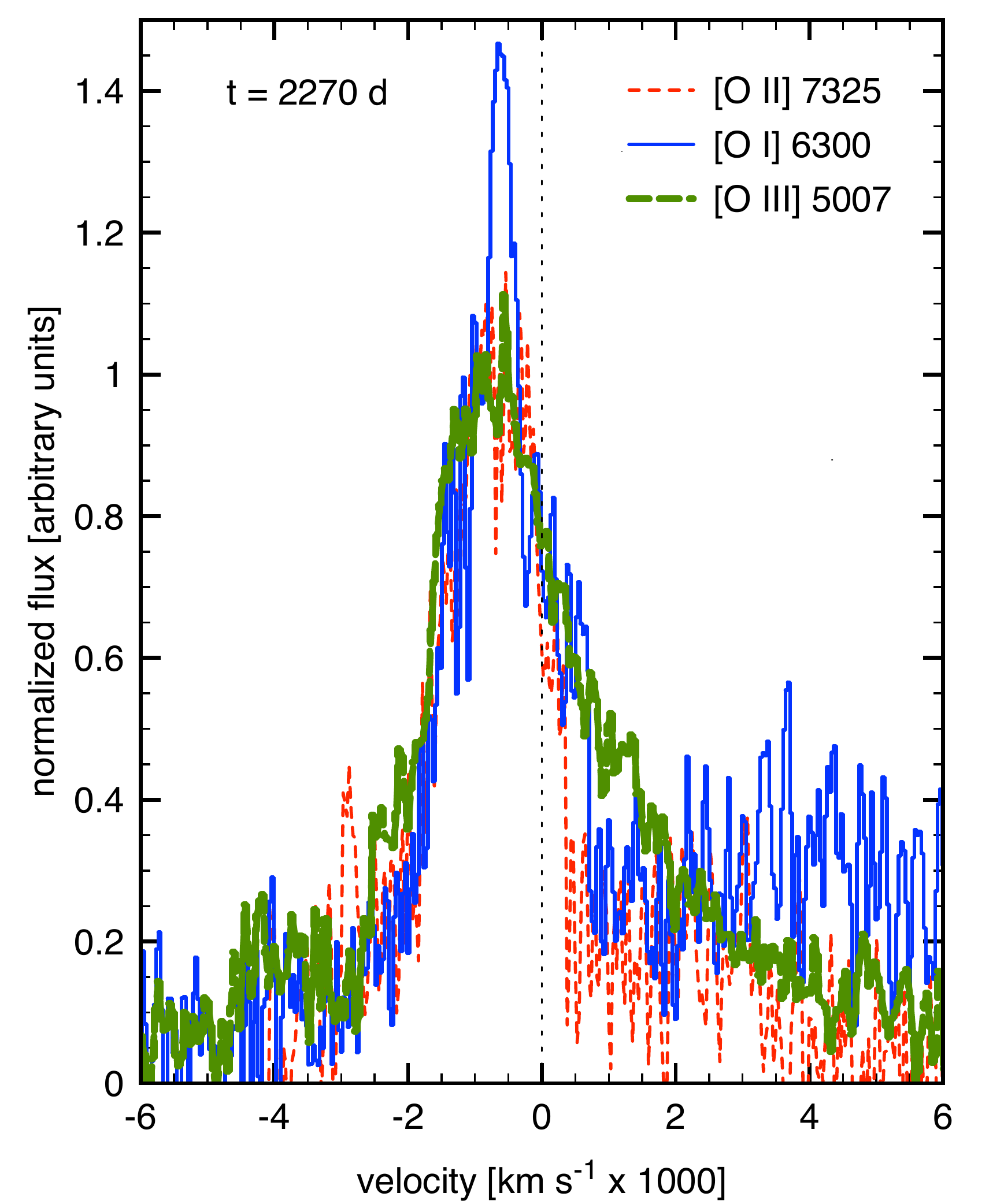}
\includegraphics[width=0.45\linewidth]{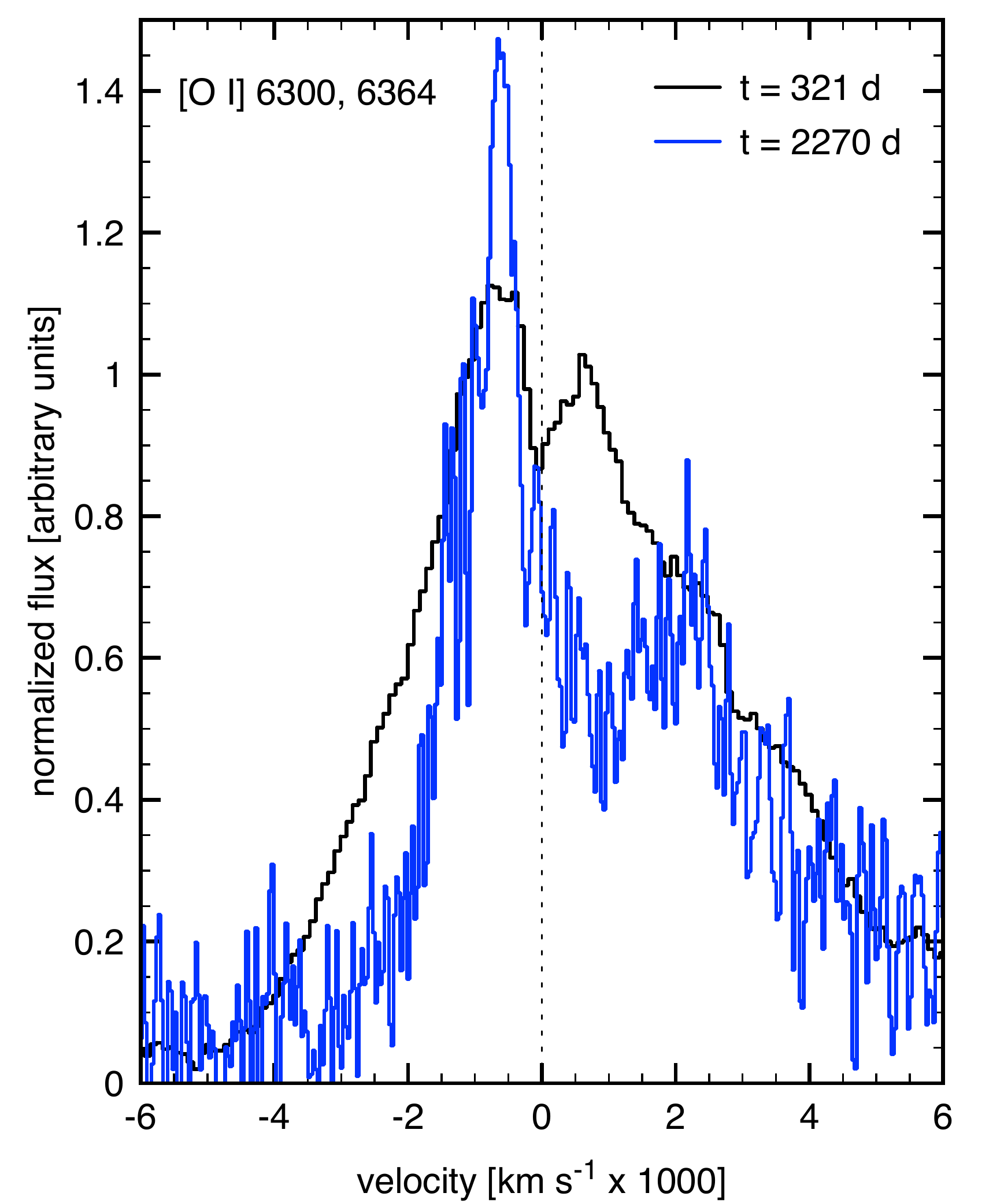}

\caption{{\it Left}: Modified [\ion{O}{2}], [\ion{O}{1}], and [\ion{O}{3}] emission line profiles of SN\,2012au. Companion doublet lines [\ion{O}{1}] $\lambda$6364 and [\ion{O}{3}] $\lambda$4959 have been modeled and subtracted from the profiles, and the [\ion{O}{2}] $\lambda\lambda$7319, 7330 blend has been treated as a single line with Doppler velocities with respect to 7325~\AA. {\it Right}: Unmodified [\ion{O}{1}] emission line profiles for days 321 and 2270.}

\label{fig:velocity}
\end{figure*}

\section{Observations} \label{sec:observations}

\subsection{Optical Spectroscopy}

A low resolution optical spectrum of SN\,2012au was obtained with the 6.5m Magellan telescope at Las Campanas Observatory on 2018 June 8. The Inamori Magellan Areal Camera and Spectrograph (IMACS; \citealt{Dressler11}) was used with the f/4 camera in combination with the Mosaic3 array of eight thinned 2K $\times$ 4K $\times$ 15-$\mu$m E2V CCDs. A 300 lines mm$^{-1}$ grating and a 0.9$^{\prime\prime}$ long slit were used. Exposures of $2 \times 3000$\,sec were obtained and averaged. Resulting spectra have an effective wavelength range of 4520-10,000 \AA, with dispersion of 0.74 \AA\ pixel$^{-1}$ and full-width-half-maximum (FWHM) resolution of 5 \AA\ (measured at 6000 \AA). The spectrum has interruptions in coverage between detector gaps. The seeing was 0.6$^{\prime\prime}$ and conditions were generally clear but not photometric.

Standard procedures to bias-correct, flat-field, and flux-calibrate the data were followed using IRAF. LA-Cosmic \citep{Dokkum01} was used to remove cosmic rays in individual images.  Some cosmetic defects introduced from hot pixels and imperfect background subtraction have been manually removed. Spectrophotometric standards Feige 56 and LTT 3864 were observed and used for absolute flux calibration \citep{Hamuy02}, which is believed to be accurate to within 30\%. The spectrum has been corrected for the redshift $z = 0.004483$ of NGC 4790 \citep{Theureau07}.

\added{\subsection{Chandra X-ray Observations}

We obtained a 20 ksec observation of SN\,2012au with the {\em Chandra X-ray Observatory} (CXO) in combination with ACIS-S on 2018 August 2 through Director's Discretionary Time (PI: Patnaude; Observation ID 21660). Data have be reduced with the CIAO software v4.10 and corresponding calibration files. In a 2$^{\prime\prime}$ radius region centered on the SN, we detect a count rate of $(3 \pm 3.5) \times 10^{-4}$\,cts\,s$^{-1}$ (0.5--10 keV), consistent with background. Thus, no X-rays are detected from SN~2012au. The neutral hydrogen column density in the direction of SN\,2012au is $3.8 \times 10^{20}$\,cm$^{-2}$ \citep{Kalberla05}. For an assumed non-thermal spectrum with index $\Gamma = 2$, we derive an unabsorbed flux limit of $2.8 \times 10^{-15}$\,erg\,s$^{-1}$\,cm$^{-2}$, corresponding to luminosity $L_{X} < 2\times 10^{38}$\,erg\,s$^{-1}$ ($0.5-10\, \rm keV$).
}

\begin{deluxetable}{cll}
\tablecolumns{3}
\tablecaption{Line fluxes and luminosities} 
\tablewidth{0pt}
\tablehead{        \colhead{Line}    	    &
                   \colhead{Flux\tablenotemark{a}}          &
                   \colhead{Luminosity\tablenotemark{b}}
}
\startdata
[\ion{O}{3}] $\lambda\lambda$4959, 5007 & $2.10 \pm 0.04$ & $1.4 \pm 0.05$ \\\
[\ion{O}{1}] $\lambda\lambda$6300, 6364 & $0.99 \pm 0.02$ & $0.65 \pm 0.02$\\\
[\ion{O}{2}] $\lambda\lambda$7319, 7330 & $0.96 \pm 0.03$ & $ 0.64 \pm 0.03$\\
\ion{O}{1} $\lambda$7774 & $0.26 \pm 0.03$ & $0.17 \pm 0.02$\\\
[\ion{S}{3}] $\lambda\lambda$9069, 9531 & $0.62 \pm 0.06$\tablenotemark{c} & $0.41 \pm 0.04$\\
\enddata
\tablenotetext{a}{In units of $10^{-15}$ erg\,s$^{-1}$\,cm$^{-2}$}
\tablenotetext{b}{In units of $10^{38}$ erg\,s$^{-1}$}
\tablenotetext{c}{Estimated from 4/3 $\times$ [\ion{S}{3}] 9531 line flux}
\label{tab:linefluxes}
\end{deluxetable}

\section{Analysis and Results} \label{sec:results}

\subsection{Spectral Properties}
\label{sec:specprops}

In Figure~\ref{fig:fullspec}, we present our day 2270 spectrum of SN\,2012au. Forbidden oxygen transitions [\ion{O}{1}] $\lambda\lambda$6300, 6364, [\ion{O}{2}] $\lambda\lambda$7319, 7330, and [\ion{O}{3}] $\lambda\lambda$4959, 5007 are clearly observed. The [\ion{S}{3}] $\lambda$9531 line is also detected but its doublet line [\ion{S}{3}] $\lambda$9069 that is intrinsically $\approx 1/3$ the strength is not recovered from noise largely associated with the subtraction of telluric lines and fringing. Broad but weak emission centered around 7775~\AA\ is also observed, which we identify with \ion{O}{1} $\lambda$7774. Emission line fluxes and luminosities are listed in Table~\ref{tab:linefluxes}.  

\begin{figure*}
\centering
\includegraphics[width=0.85\linewidth]{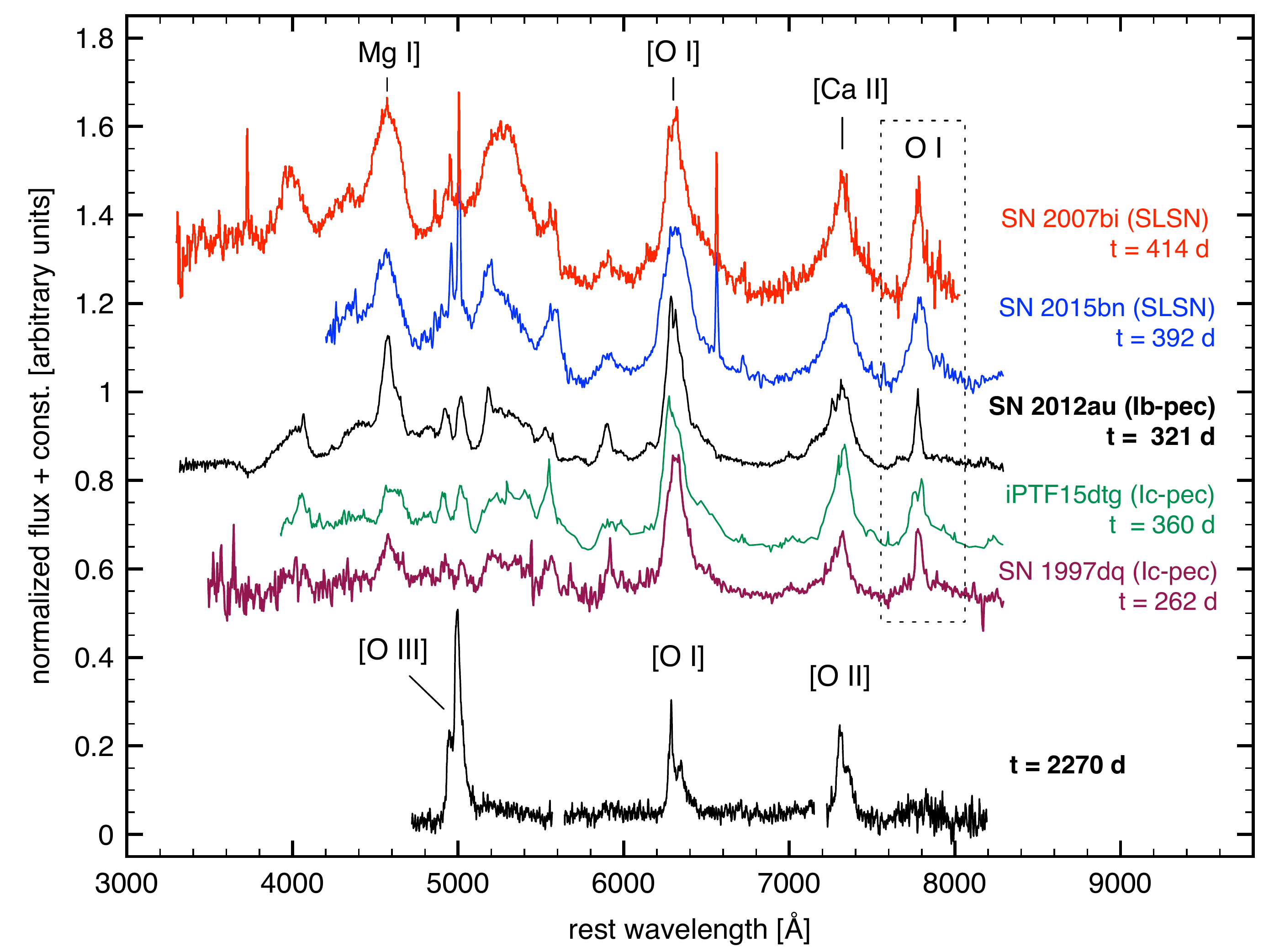}

\caption{Earlier nebular spectrum ($t \approx 1$ yr) of SN 2012au \citep{Milisavljevic13} compared to our $t = 6.2$\,yr spectrum. Also shown is a subset of related SLSNe (e.g., SNe 2007bi and 2015bn; \citealt{Gal-Yam09} and \citealt{Nicholl16}), and SNe\,Ic (SN\,1997dq and iPTF15dtg; \citealt{Matheson01} and \citealt{Taddia18}). The \ion{O}{1} $\lambda$7774 line, which is a defining feature of the related objects, has faded. All events may evolve to a similar late-time emission phase.}

\label{fig:speccompare}
\end{figure*}

All forbidden line profiles exhibit a clear asymmetry towards blueshifted wavelengths, peaking at $-700 \pm 50$\,\kms\ (Fig.~\ref{fig:velocity}). We measure the expansion velocity of [\ion{O}{3}], [\ion{O}{1}], [\ion{O}{2}], and [\ion{S}{3}] using the half-width-zero-intensity (HWZI) shortward of 4959~\AA, 6300~\AA, 7325~\AA, and 9531~\AA, respectively. All measurements extend to approximately $2300 \pm 100$ \kms.  The weakly detected \ion{O}{1}~7774 has no clear peak in emission and does not appear to have the same emission line profile as the forbidden transitions.

Excess flux is observed between 5100-5400 \AA. This could potentially be the remnant of the ``plateau'' of emission observed in the day 321 spectrum (\citealt{Milisavljevic13}; Fig.~\ref{fig:speccompare}). However, contaminating flux from the host galaxy cannot be ruled out. The fall-off in flux below 4800~\AA\ is attributed to rapid loss of instrument sensitivity at these shorter wavelengths.

No narrow (FWHM $<200$ \kms) features are observed in the spectrum. Particular attention was made to possible emission in the region of H$\alpha$, which would be indicative of interaction with H-rich CSM. \added{The 2D spectrum shows narrow emission lines of [\ion{O}{1}] 6300, 6364, [\ion{N}{2}] 6548, 6583, and H$\alpha$ extended spatially across the slit but nothing conspicuous and specific to the SN location (Fig.~\ref{fig:fullspec}, bottom). After careful subtraction of the local background} we estimate an upper limit of H$\alpha$ emission to be $< 4.1 \times 10^{-17}$\,erg\,s$^{-1}$\,cm$^{-2}$.   

\subsection{Emission line diagnostics}
\label{sec:diagnostics}

The emission line intensity ratios of our spectra can be used with atomic rates from CHIANTI \citep{DelZanna15} to constrain properties of the ejecta. The intensity ratio \ion{O}{1}/[\ion{O}{2}] is a sensitive temperature diagnostic and the measured ratio of $\approx 0.27$ indicates a temperature $T \approx 5000$ K.  However, this should be viewed as an average temperature of the O$^+$ zone. Because the line profile of \ion{O}{1} does not closely follow those of the other lines (section \ref{sec:specprops}), some emission may originate from a different region of the ejecta and include cool gas that does not emit in the collisionally excited lines.

The [\ion{O}{3}]/[\ion{S}{3}] ratio reflects a higher temperature of the O$^{++}$/S$^{++}$ zone.  If we assume that (S$^{++}$/S) = (O$^{++}$/O), then the measured ratio of [S\,III]/[O\,III] = 0.146 can be used in the expression
\begin{equation*}
{\rm [S\,III]/[O\,III]} = 5.98 \times {\rm [S/O]} \times  \exp{(12800/T)}
\end{equation*}
to give
\begin{equation*}
{\rm S/O} = 0.024 \times \exp{(-12800/T)}.
\end{equation*}
The assumption of equal ionization fractions for S and O is plausible, but it potentially introduces an uncertainty of a factor of 2. 

Notably, [\ion{S}{3}] $\lambda$9531 line emission is observed, but [\ion{S}{2}] $\lambda\lambda$6717, 6731 is absent.  This suggests electron densities above $10^4~\rm cm^{-3}$.  Using the ratio of S/O from the equation above at $10^4$\,K, assuming (S$^+$/S) = (O$^+$/O) in the singly ionized zone, and that $\log{T}$ is at least 3.7, then the ratio of [S II]/[O II] as a function of density can be determined. Estimating the upper limit to emission centered around the [\ion{S}{2}] $\lambda\lambda$6716,6731 lines to be $\approx 0.2 \times I$([\ion{O}{2}]), we find that $\log{T}=3.8$ and $\log{n} \ga 6.0$.  

We thus conclude a sulfur to oxygen ratio $\sim 0.01$ and a density of $\log{n} \ga 6.0$. The high density is likely associated with significant clumping of ejecta, because a uniform sphere at that density with the radius $R \sim 4.3 \times 10^{16}$\,cm given by the time since explosion and observed expansion velocity would produce far more than the observed luminosity. The line profiles of all the forbidden oxygen lines are roughly the same (Fig.~\ref{fig:velocity}), indicating that they are likely co-located, but with O$^{++}$ associated with a lower density, higher temperature zone.

\section{Discussion} \label{sec:discussion}

Our $t = 6.2$ yr spectrum of SN\,2012au is markedly different from the last published spectrum at $t \sim 1$\,yr (\citealt{Milisavljevic13}; Fig.~\ref{fig:speccompare}). At that time SN\,2012au exhibited [\ion{O}{1}], [\ion{Ca}{2}], and \ion{Mg}{1}] emissions typical of stripped-envelope core-collapse SNe, and unusually strong emissions from \ion{Ca}{2}~H\&K, \ion{Na}{1}~D, and \ion{O}{1}~7774. It also showed persistent P-Cyg absorptions attributable in part to \ion{Fe}{2} at 2000 \kms. None of these features are observed in the new spectrum, with the exception of [\ion{O}{1}] that has a radically different emission line profile (Fig.~\ref{fig:velocity}) and the weakly detected \ion{O}{1}.

Figure~\ref{fig:speccompare} also shows examples of objects that share late-time emission properties of SN 2012au.  The recently reported optical spectrum of iPTF15dtg \citep{Taddia18} that we identify as a member of this grouping is included as well. \citet{Milisavljevic13} noted that asymmetries between the emission line profiles of ions in these objects are consistent with moderately aspherical explosions. They also highlighted the \ion{O}{1}~$\lambda$7774 line of width $\approx 2000$ \kms\ as being indicative of a jetted explosion and a defining feature of these objects. \citet{Nicholl16} later interpreted the \ion{O}{1} feature to be the signature of heating of a shell by a central engine. 

Below we discuss possible excitation mechanisms for the late-time spectrum of SN\,2012au and utilize multi-wavelength data that further constrain properties of the ejecta and CSM.

\subsection{Radioactivity and SN--CSM interaction}

Typically, late-time spectra of SNe~I at stages reaching one year are still powered by radioactive $^{56}$Co in the $^{56}$Ni$\rightarrow$$^{56}$Co$\rightarrow$$^{56}$Fe decay chain. However, radioactivity is not a plausible heating source for the oxygen-rich ejecta at $t = 6.2$~yr considering $\approx 0.3$\,M$_{\odot}$ of $^{56}$Ni was produced in SN\,2012au \citep{Milisavljevic13,Takaki13} and $^{56}$Co has a half-life of 77.3 days \citep{Alburger89}. Furthermore, radioactivity is typically associated with neutral and singly ionized lines \citep[see, e.g.,][]{Matheson01}, and yet our 6.2\,yr spectrum exhibits a larger range of ionization levels.

\begin{figure}
\centering
\includegraphics[width=\linewidth]{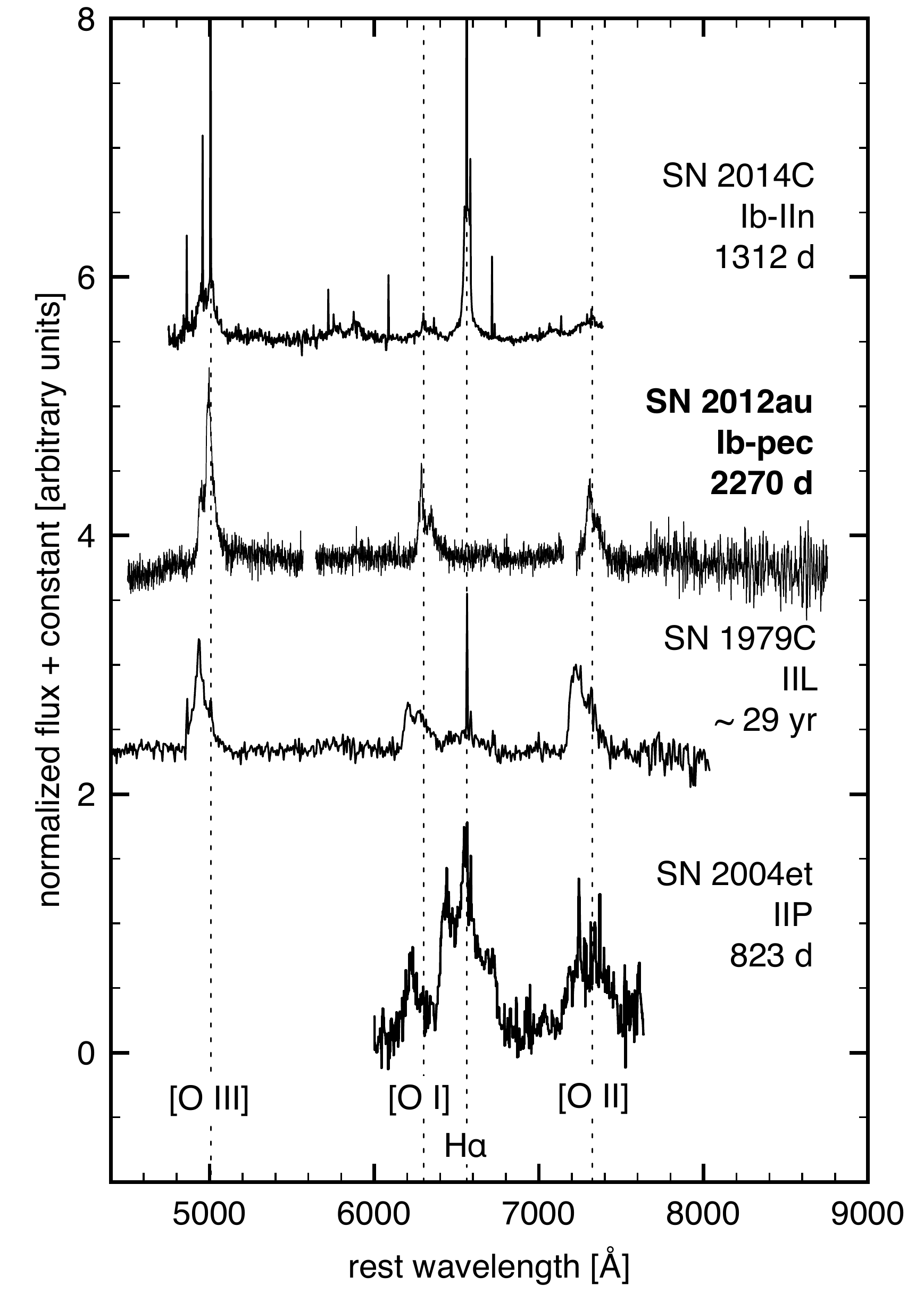}

\caption{With the exception of SN\,2012au, optical spectra of SNe observed at epochs $t > 3$\,yr show hydrogen lines and evidence for SN--CSM interaction. Examples shown here are SN\,2014C \citep{Mauerhan18}, SN\,1979C \citep{Milisavljevic09}, and SN\,2004et \citep{Kotak09}.}

\label{fig:latecompare}
\end{figure}

SN--CSM interaction is the most common late-time emission mechanism for objects observed $> 3$ yr post-explosion (see \citealt{Milisavljevic12} and \citealt{CF17} and references therein). Optical emission principally originates from a reverse shock that propagates upstream into outward expanding ejecta that gets heated and ionized. Such late-time detections all have clear signatures of interaction with an H-rich CSM including development of narrow emission lines and/or high velocity H-rich ejecta (\citealt{MF17}; Fig.~\ref{fig:latecompare}). SNe Ib,c examples include SN\,2001em \citep{Chugai06}, SN\,2014C \citep{Milisavljevic15}, and SN\,2004dk \citep{Mauerhan18}. A few SLSNe-I have exhibited hydrogen emission several hundred days after explosion \citep{Yan17}. This contrasts with SN\,2012au, which shows no spectral features indicative of SN--CSM interaction. 

\added{If SN\,2012au was indeed interacting with H-rich material, then H Balmer line emission would be expected. Certain H-rich CSM distributions could potentially inhibit optical emission at the time of observation, but such configurations are considered to be unlikely.} It is also possible that SN\,2012au is interacting with an H-poor environment. This scenario has been discussed in the context of SLSNe-I \citep{Chatzopoulos12}, but the detailed spectroscopic properties predicted for helium/carbon/oxygen-rich CSM interaction at these extremely late epochs is poorly explored. 

%Alternatively, it could be that the forward shock has run through all or nearly all of the CSM shell.  In that case, there is no photoionized CSM that would produce the optical lines, since all the gas is very hot.

\subsection{Pulsar Wind Nebula} \label{sec:pwn}

Another late-time emission mechanism is pulsar interaction with expanding SN gas. A PWN is generated by the spin-down power of a central pulsar. In this scenario, photoionization of the inner regions of the expanding shell of ejecta can be the dominant source of optical line emission, especially at early times ($< 10$ yr) when the expanding ejecta absorb much of the PWN-emitting ionizing radiation.

The \citet{CF92} model is based on the young Crab Nebula and has been widely used for pulsar and magnetar nebulae (e.g., \citealt{Kasen10}). The freely expanding SN density profile is approximated by an inner, flatter power law density profile surrounded by a steeper one.  The inner profile has index $m=1$ and the outer $n=9$ \citep{RS89,MM99}.  For the SN\,2012au parameters $E_k \sim 10^{52}$\,erg and $M_{ej} \approx 4\,M_{\odot}$, the transition between these power laws occurs at 18,000 \kms. Thus the SN\,2012au PWN is well within the inner, flatter region of the freely expanding gas.  From equ. (2.11) of \cite{CF92}, the velocity $V$ of the PWN is  
\begin{equation}
V=289 \dot{E}^{0.25}_{39} E_{k51}^{0.25} M_{ej1}^{-0.5} t_{yr}^{0.25} \quad {\rm km~s^{-1}},
\label{vel}
\end{equation}
where $\dot{E}_{39}$ is in units of $10^{39}$ erg s$^{-1}$, $E_{k51}$ in $10^{51}$ ergs, ${M_{ej1}}$ in 10 $M_{\odot}$, and $t_{yr}$ in years.
With $V=2300$\,km\,s$^{-1}$ (assuming the emission is coming from close to the shell), $E_{k51}=10$, ${M_{ej1}}=0.4$, and $t_{yr}=6.2$, we have $\dot{E}=10^{40}$ erg s$^{-1}$. The Crab pulsar currently has $\dot{E} \approx 4\times 10^{38}$\,erg\,s$^{-1}$ \citep[e.g.,][]{condon16} and extrapolating back with constant braking index gives an initial $\dot{E} \approx 4 \times 10^{39}$\,erg\,s$^{-1}$, quite close to what is needed for SN\,2012au.  The Crab magnetic field is $B \approx 4 \times 10^{12}$\,G \citep[e.g.,][]{condon16}, well below the magnetar field $B>10^{14}$ G.  Hence, the pulsar in SN\,2012au is potentially more Crab-like than magnetar-like.

The \citet{CF92} model includes line estimates for an O zone.  It predicts an [\ion{O}{3}] luminosity at 1500 days to be $0.46 \times 10^{38}$\,erg\,s$^{-1}$, which is close to the $1.4 \times 10^{38}$\,erg\,s$^{-1}$ observed in SN\,2012au. [\ion{O}{1}] is weak because a thermal instability cools the gas, but somewhat different parameters could increase [\ion{O}{1}].  [\ion{O}{2}] is weak because of low temperature and high density, but that could also change.  The high energy of SN\,2012au should result in a considerably lower density that the PWN is moving into than in the Crab.

\citet{CF92} assumed that the swept up shell would be broken up by instabilities so the ionizing radiation from the PWN photoionizes freely expanding gas ahead of the shock front.  In that case, there are different layers of ionized and neutral O in the O zone.  However, the similar line profiles for [\ion{O}{1}], [\ion{O}{2}], and [\ion{O}{3}] (Fig.~\ref{fig:velocity}) do not support this picture.

A radiative shock wave driven into the freely expanding ejecta by the PWN is another possible excitation mechanism.  However, the efficiency of this process is low, $ L \sim 0.015 \times \dot{E}$.  Also, one would expect the characteristic boxy line profile for shell emission, which is not observed.

The most likely scenario seems to be photoionization of O zone gas that has been shocked by the high pressure PWN and subjected to instabilities. In the Crab, the photoionization is of the H/He zone.  Both SN\,2012au and the Crab show fairly centrally peaked line profiles, as would be expected if Rayleigh-Taylor instabilities mix gas to the central region. \citet{BC17} performed a simulation of this process (see Fig.\ 3 of that paper). Dense gas is in both Rayleigh-Taylor fingers and an outer shocked shell. Photoionization layers are anticipated, but on a small scale, so the line profiles of different ions will be similar.  \added{The compression in the PWN-driven shock also leads to the high density deduced from line diagnostics (section \ref{sec:diagnostics}).}

A prediction of the above pulsar model is broadening of the emission lines with time because of the acceleration of the pulsar bubble. Following \citet{CF92}, one would expect roughly steady pulsar power leading to $R \propto t^{1.25}$ for the swept up shell or velocity $V \propto t^{0.25}$. Specifically for SN\,2012au, we  anticipate
\begin{equation*}
\delta V/V = 0.25\,\delta t/t = 0.25 (1~{\rm yr}/ 6.2 ~{\rm yr}) = 0.04
\end{equation*}
in one year, or a rate of increase in the emission line velocity width of $\approx 4$\%\,yr$^{-1}$.

In the above model, it is assumed that the initial pulsar spindown timescale is much greater than the age. If the spindown timescale is much less than the age, the pulsar input occurs early and the shell tends to a constant velocity; the shell comoves with the freely expanding gas and $V=R/t$.  The kinetic energy of the shell depends on the mass, which can be found from the freely expanding density profile, and velocity $V$, yielding $E_k=2.5\times 10^{48}$ ergs.  The energy of freely expanding gas is $1.3\times 10^{48}$ ergs, which implies that the initial rotational energy of the pulsar, $E_0$, was $1.3\times 10^{48}$ ergs, corresponding to a rotation period of 0.13 s.  This is longer than the ms periods found in the magnetar models for SLSNe and is due to the $V^4$ dependence of $E_k$ (eqn. (\ref{vel})); line widths in the magnetar model approach $\sim 12,000$ km s$^{-1}$ and remain constant.

We favor the long spindown model because it implies a current pulsar power that is roughly consistent with that needed to produce the observed luminosity.  A prediction of this model is increasing PWN velocities, as opposed to constant velocities in the short spindown case. Deceleration is not expected in a PWN model.  

\added{\section{X-ray emission from the PWN}
\label{sec:x-ray}

We considered the viability of detecting X-rays from the candidate PWN. Assuming a dipole model for the central source, with a spin period of 1 ms, the spindown luminosity is given as

\begin{equation*}
L_p \approx 2 \times 10^{42} B_{14}^2 (t/\mathrm{yr})^{-2} \, \mathrm{erg\,s^{-1}}\, ,
\end{equation*}

\noindent where $B_{14}$ is the magnetic field in units of 10$^{14}$ G. At the time of our {\em CXO} observation ($t = 2325$ d), $L_p$ $\lesssim$ 10$^{41}$ erg\,s$^{-1}$. If only 1\% of that is converted into X-rays, the expected luminosity of the central source is $\sim 10^{39}$\,erg\,s$^{-1}$. This would be the most favorable situation, and a less energetic PWN would have a lower luminosity. 

We can estimate the optical depth to 1 keV X-rays as

\begin{equation*}
\tau = (\sigma /16\,m_p) \times 3/(4\pi) * M_{ej}/(v_{ej} t)^{2} \, ,
\end{equation*}

\noindent where $\sigma$ is the cross section for 1 keV X-rays, $m_p$ is the proton mass, $v_{ej}$ is the initial ejecta velocity, and $t$ is the age in years. Assuming a cross section of 10$^{-19}$\,cm$^{2}$, appropriate for oxygen-rich ejecta, and $v_{ej}$ upwards of $\sim 3 \times 10^4$\, \kms, the optical depth to 1 keV X-rays is $\gg 1$. Thus, the {\em CXO} non-detection is consistent with our estimate that the candidate PWN of SN\,2012au is presently optically thick to 1\,keV X-rays.

Notably, SN--CSM interaction (see \citealt{CF17}) is a possible alternative source of late-time X-ray emission. However, adopting parameters estimated in \citet{Kamble14} (progenitor mass loss rate $\dot{M}$ $\approx$ 4 $\times$ 10$^{-6}$ M$_{\odot}$\,yr$^{-1}$, with a wind speed of $v_w = 1000$ \kms\ extending out to a radius of at least 10$^{17}$ cm), the expected thermal X-ray emission associated SN--CSM interaction would only be $\lesssim$ 10$^{32}$ erg\,s$^{-1}$. This translates to a 0.5--10 keV X-ray flux of $\sim$ 10$^{-20}$ erg\,cm$^{-2}$\,s$^{-1}$, which is well below our detection limit.
}

\section{Conclusions} \label{sec:conclusions}

We have presented optical \added{and X-ray} observations of the energetic, slow-evolving, Type Ib  SN\,2012au obtained $> 6$~yr post-explosion that provide direct evidence of a newly-formed PWN exciting O-rich ejecta. Our findings support the notion that SN\,2012au and a subset of SLSNe, GRB-SNe, and SNe\,Ib/c have been relatedly influenced by magnetized compact objects on a wide range of energy scales. It remains unclear what key aspects of the progenitor systems unite these SNe that span absolute magnitudes of $-22 < M_B < -17$.  

We anticipate that, like SN\,2012au, SNe harboring influential pulsar/magnetar wind nebulae will evolve into a late-time phase dominated by forbidden oxygen transitions. Furthermore, we predict that optical emission line widths should remain constant or broaden upwards of a few per cent per year due to acceleration of ejecta by the pulsar/magnetar bubble. In the specific case of SN\,2012au, we estimate velocity broadening at the rate of 4\%\,yr$^{-1}$. Measurements are potentially achievable with return visits from 8-10-m class telescopes using $R \ga 10000$ spectroscopy over the next several years, although best opportunities for distant events await the arrival of next-generation Extremely Large Telescopes.

\acknowledgments

We thank the referee for a very helpful reading of the paper and suggestions that significantly improved its content and presentation. This paper includes data gathered with the 6.5 meter Magellan Telescopes located at Las Campanas Observatory, Chile. D.J.P.\ acknowledges support from NASA Contract NAS8-03060. We thank the {\em Chandra X-ray Observatory} for the Director's Discretionary Time which made possible the X-ray observations cited in this paper. Data for SNe 1997dq 2007bi, and 2015bn were retrieved from the Weizmann interactive supernova data repository (\url{http://wiserep.weizmann.ac.il}). J.\ Mauerhan kindly shared spectra of SN\,2014C. R.A.C.\ acknowledges support from NSF grant AST-1814910.

\vspace{5mm}
{\it Facilities:} Magellan: Baade (IMACS), CXO

\end{document}